\documentclass[aps,pra,preprint,superscriptaddress,longbibliography]{revtex4-2}

\usepackage{amsmath}
\usepackage{amsthm}
\usepackage{amsfonts}
\usepackage{amssymb}
\usepackage{xcolor,graphicx}
\usepackage[pdftex]{hyperref} 
\usepackage{longtable}
\usepackage{array}
\usepackage{dsfont}
\usepackage{tabularx}
\usepackage{stix}

\definecolor{cz0}{HTML}{ffffff}
\definecolor{cz1}{HTML}{6665ff}
\definecolor{cz2}{HTML}{d1d1d1}
\definecolor{cz3}{HTML}{ffff66}
\definecolor{cz4}{HTML}{ff6666}
\definecolor{cz5}{HTML}{9d9ce8}
\definecolor{cz6}{HTML}{66ff66}
\definecolor{cz7}{HTML}{b166b1}
\definecolor{cz8}{HTML}{e7e89c}
\definecolor{cz9}{HTML}{e89d9b}
\definecolor{cz10}{HTML}{ffb166}
\definecolor{cz11}{HTML}{9ee89d}
\definecolor{cz12}{HTML}{c29cc4}
\definecolor{cz13}{HTML}{c2a386}
\definecolor{cz14}{HTML}{e8c29d}
\definecolor{cz15}{HTML}{cccccc}

\begin{document}
	
	\title{Routes to Chaos in Class-B laser Dimer Necklaces}
	
	\author{J. Diaz-Avila}
	\email[e-mail: ]{jonathandiazavila2@gmail.com}
	\affiliation{}	

	\author{A.~A. {Macias Infante}}
	\affiliation{}	

	\author{B.~R. Jaramillo-\'Avila}
    \email[e-mail: ]{bjaramillo@cicese.mx}
	\affiliation{Monterrey Unit, Centro de Investigacion Cientifica y de Educacion Superior de Ensenada Apodaca, N.L., C.P. 66629, Mexico}
 
	\author{B.~M. Rodr\'iguez-Lara}
    \email[e-mail: ]{blas.rodriguez@gmail.com}
	\affiliation{Universidad Polit\'ecnica Metropolitana de Hidalgo, Tolcayuca, Hidalgo 43860, Mexico.}	
	
	\date{\today}
	
    \begin{abstract}
    Linear $\mathcal{PT}$-symmetric dimer necklaces use cyclic coupling to organize phase-locked modes, while class-B laser--resonator dimers use carrier inversion to destabilize gain--loss dynamics and produce routes to chaos.
    We study a cyclic necklace of class-B laser--resonator dimers to determine how inter-dimer coupling and cyclic phase constraints modify carrier-mediated instabilities.
    The carrier inversion turns the necklace into an amplitude--phase--carrier system whose active-site gain controls intensities, while the linewidth-enhancement factor shifts phases.
    We derive the real dynamical equations, identify uniform fixed points above threshold, and classify their phase configurations under cyclic closure.
    Even necklaces support full synchronization, a uniform $\pi$ shift, and two alternating phase configurations; odd necklaces support only the first two.
    Linear stability of these configurations partitions the coupling plane into sixteen possible regions, including one region with no stable uniform fixed point.
    Long-time simulations for representative two- and three-dimer necklaces show stable lasing, multistability, and irregular lasing controlled by pump ratio and coupling strengths.
    Increasing the pump ratio expands stable domains and suppresses the region without stable uniform fixed points.
    \end{abstract}
	
	
	\maketitle
	\newpage

\section{Introduction} 
\label{sec:Sec1}

Integrated photonic devices support classical and quantum information processing~\cite{Brunner2020,Pelucchi2022,PsiQuantum2025}, optical communications~\cite{Chovan2018,Glick2023}, and sensing~\cite{Arafin2018}. 
Their current development combines material platforms, active devices, and circuit architectures~\cite{Zhou2023}. 
Evanescent coupling turns integrated elements into arrays where synchronization produces coherent emission~\cite{Gourley1991,Garcia1999}, dynamical instabilities produce deterministic chaos~\cite{Winful1990,Winful1992,Thornburg1997,Arroyo2012,Adams2022}, gain--loss contrast produces spontaneous $\mathcal{PT}$-symmetry breaking~\cite{ElGanainy2018}, engineered band structure produces topological edge transport~\cite{Bandres2018,Harari2018,Longhi2018,Parto2018,Zhao2018}, and path interference enhances or suppresses decay~\cite{JaramilloAvila2020b,Longhi2020}.

Linear gain--loss imbalance gives a controlled route to organize collective modes in those arrays.
Parity--time-symmetric non-Hermitian Hamiltonians can support real spectra below a symmetry-breaking threshold~\cite{Bender1998}, and photonic systems realize the same structure with coupled elements carrying balanced gain and loss~\cite{ElGanainy2007}. 
Waveguides~\cite{Ruter2010}, directional couplers~\cite{Ruter2010}, waveguide arrays~\cite{HuertaMorales2016}, and microring resonators~\cite{Hodaei2014,Hodaei2016,Ren2018} provide direct control of gain--loss contrast and coupling strength. 
In the linear regime, this control moves the spectrum from real eigenvalues in the $\mathcal{PT}$-symmetric phase to complex-conjugate pairs in the broken phase through an exceptional point, where eigenvalues and eigenvectors coalesce and the evolution contains a secular term~\cite{Zyablovsky2014a,Zyablovsky2014b}.

Optical $\mathcal{PT}$ analogs usually operate in class-A or effectively linear-gain regimes, where carrier and polarization dynamics are not independent variables~\cite{Hodaei2014,Hodaei2016,Baili2009}.
Class-B semiconductor lasers give a controlled route beyond that limit by retaining the carrier inversion while the polarization follows the field~\cite{Ohtsubo2017}.
This carrier feedback replaces linear gain--loss evolution by nonlinear amplitude--phase dynamics with relaxation oscillations, multistability, and chaos in coupled laser systems~\cite{Winful1990,Winful1992,Thornburg1997,Adams2022}.

A class-B laser coupled to a neutral resonator gives the minimal nonlinear gain--loss dimer~\cite{GuemesFrese2022}. 
Above threshold, the isolated laser reaches a steady lasing state, while coherent coupling to the neutral resonator returns the field to the active site and destabilizes that state. 
This carrier-mediated feedback produces fixed points, limit cycles, multistability, and routes to chaos without delayed feedback or external modulation~\cite{GuemesFrese2022}. 
An open problem is how these instabilities change when the dimer becomes a cyclic array with phase closure. 
The linear $\mathcal{PT}$-symmetric necklace shows that cyclic coupling can copy single-dimer dynamics across the array through symmetry-protected modes~\cite{NodalStevens2018}. 
A class-B necklace places that cyclic constraint on nonlinear amplitude--phase--carrier dynamics, where intra-dimer laser--resonator coupling, inter-dimer coupling, and active-site carrier inversion act together.

Here, we study a class-B dimer necklace, Sec.~\ref{sec:Sec2}, by combining the cyclic $\mathcal{PT}$-symmetric necklace with the class-B laser--resonator dimer.
Our construction gives a real amplitude--phase--carrier model where intra-dimer coupling, inter-dimer coupling, and carrier inversion act on the same variables.
We use the uniform fixed points to impose cyclic phase closure, separate the even- and odd-necklace phase configurations, and partition the coupling plane by linear stability, Sec.~\ref{sec:Sec3}.
Representative two-dimer and three-dimer necklaces show how pump ratio and coupling strengths organize stable lasing, multistability, and chaotic lasing.
We conclude, Sec.~\ref{sec:Sec4}, by identifying cyclic phase closure as the constraint that separates the even- and odd-necklace routes to carrier-mediated nonlinear dynamics.

\section{Class-B Dimer Necklace} 
\label{sec:Sec2}

We model a class-B necklace by adding carrier-inversion dynamics to the active sites of a linear $\mathcal{PT}$-symmetric dimer necklace.

The ideal $\mathcal{PT}$-symmetric dimer~\cite{ElGanainy2007},
\begin{align}
    \begin{aligned}
        i \dot{E}_{1}(t) =&~ \left( \omega + i \gamma \right) E_{1}(t) + g_{d} E_{2}(t), \\
        i \dot{E}_{2}(t) =&~ \left( \omega - i \gamma \right) E_{2}(t) + g_{d} E_{1}(t),
    \end{aligned}
\end{align}
describes two coupled complex field amplitudes $E_{1}(t)$ and $E_{2}(t)$ with common resonant frequency $\omega$, gain--loss coefficient $\gamma$, and coupling strength $g_{d}$, all with units of frequency.
This dimer has three spectral regimes.
For $\gamma < g_{d}$, the eigenvalues are real and the field amplitudes oscillate.
At $\gamma = g_{d}$, the exceptional point coalesces the two eigenvalues and eigenvectors.
For $\gamma > g_{d}$, the eigenvalues form a complex-conjugate pair and the amplitudes show exponential growth and decay~\cite{RodriguezLara2015}.
Exceptional-point spectral splitting has been proposed for enhanced sensing~\cite{Chen2016}, while coupled class-A lasers approximate this linear gain--loss setting when carrier and polarization relaxation are fast~\cite{Baili2009}.

The linear dimer necklace couples $N$ ideal $\mathcal{PT}$-symmetric dimers in a cyclic array~\cite{NodalStevens2018},
\begin{align}
    \begin{aligned}
        i \dot{E}_{2k}(t) =&~ \left( \omega + i \gamma \right) E_{2k}(t) + g_{d} E_{2k+1}(t) + g_{N} E_{2k-1}(t), \\
        i \dot{E}_{2k+1}(t) =&~ \left( \omega - i \gamma \right) E_{2k+1}(t) + g_{d} E_{2k}(t) + g_{N} E_{2k+2}(t),
    \end{aligned}
\end{align}
with $k = 0, \ldots, N - 1$ and all indices modulo $2N$.
The coupling $g_{d}$ links the active and passive sites inside each dimer, while $g_{N}$ closes the ring through inter-dimer links.
Local $\mathcal{PT}$ symmetry remains dimerwise, and cyclic symmetry decomposes the necklace into $N$ effective dimers with cyclic-mode-dependent couplings.
For homogeneous couplings, even necklaces always contain a broken-symmetry block, while unequal intra- and inter-dimer couplings can restore the $\mathcal{PT}$-symmetric phase.
The same cyclic symmetry allows input phase patterns to select an effective dimer and replicate its dynamics across the necklace.

A class-B laser--resonator dimer replaces the active linear-gain site with a class-B laser~\cite{GuemesFrese2022},
\begin{align}
    \begin{aligned}
        i \dot{E}_{1}(t) =&~ \left[ \omega + \left( \frac{i + \alpha}{2} \right) \left\{ \sigma \left[ n(t) - 1 \right] - \frac{1}{\tau_{p}} \right\} \right] E_{1}(t) + g_{d} E_{2}(t), \\
        i \dot{E}_{2}(t) =&~ \omega E_{2}(t) + g_{d} E_{1}(t), \\
        \dot{n}(t) =&~ R - \frac{n(t)}{\tau_{s}} - \frac{2 \left[ n(t) - 1 \right]}{\tau_{s}} |E_{1}(t)|^{2}.
    \end{aligned}
\end{align}
The fields $E_{1}(t)$ and $E_{2}(t)$ describe a class-B laser coupled to a neutral resonator, and $n(t)$ is the carrier inversion normalized to transparency.
The linewidth-enhancement factor, differential gain, photon lifetime, carrier lifetime, and pump rate are $\alpha$, $\sigma$, $\tau_{p}$, $\tau_{s}$, and $R$, respectively~\cite{Arecchi1984,Wieczorek2005,Baili2009}.
This dimer has two zero-detuning fixed points above threshold, with equal amplitudes and phase difference $0$ or $\pi$~\cite{GuemesFrese2022}.
Their stability partitions the pump--coupling plane into stable lasing, bistable lasing, and chaotic or nonstationary regions in the class-B hierarchy $\tau_{s} \gg \tau_{p} \gg \sigma^{-1}$~\cite{Arecchi1984,Baili2009,GuemesFrese2022}.

\begin{figure}
    \centering
    \includegraphics[scale=1]{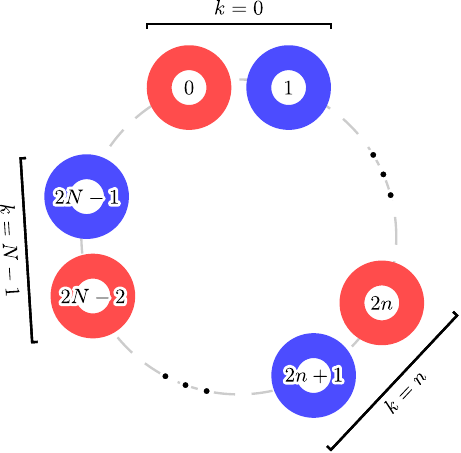}
    \caption{Necklace of class-B laser dimers. 
    Red sites correspond to class-B lasers, and blue sites correspond to neutral resonators.}
    \label{fig:Fig1}
\end{figure}

Our class-B dimer necklace places a class-B laser at each even site $2k$ and a neutral resonator at each odd site $2k+1$,
\begin{align}
    \begin{aligned}
        i \dot{E}_{2k}(t) =&~ \left[ \omega + \left( \frac{i + \alpha}{2} \right) G_{2k}(t) \right] E_{2k}(t)  + g_{d} E_{2k+1}(t) + g_{N} E_{2k-1}(t), \\
        i \dot{E}_{2k+1}(t) =&~ \omega E_{2k+1}(t) + g_{d} E_{2k}(t) + g_{N} E_{2k+2}(t), \\
        \dot{n}_{2k}(t) =&~ R - \frac{n_{2k}(t)}{\tau_{s}} - \frac{2 \left[ n_{2k}(t) - 1 \right]}{\tau_{s}} |E_{2k}(t)|^{2},
    \end{aligned}
\end{align}
with $k = 0, \ldots, N - 1$ and all indices modulo $2N$.
The active-site net gain,
\begin{align}
    G_{2k}(t) =&~ \sigma \left[ n_{2k}(t) - 1 \right] - \frac{1}{\tau_{p}},
\end{align}
couples carrier inversion to the active-site amplitude and phase through the factor $(i + \alpha) / 2$.
The coupling $g_{d}$ links the laser and resonator inside each dimer, while $g_{N}$ links neighboring dimers and closes the necklace.

We use polar fields and nearest-neighbor phase differences to remove the global optical phase and keep the cyclic phase constraints explicit,
\begin{align}
    \begin{aligned}
        E_{k}(t) =&~ \sqrt{A_{k}(t)} \, e^{i\phi_{k}(t)}, \\
        \Delta\phi_{k}(t) =&~ \phi_{k+1}(t) - \phi_{k}(t).
    \end{aligned}
\end{align}
This representation leaves $2N$ amplitudes, $2N - 1$ independent phase differences, and $N$ carrier inversions,
\begin{align}
    \begin{aligned}
        \dot{A}_{2k}(t) =&~ 2g_{d}\sqrt{A_{2k}(t)A_{2k+1}(t)} \, \sin \Delta\phi_{2k}(t)  - 2g_{N}\sqrt{A_{2k}(t)A_{2k-1}(t)} \, \sin \Delta\phi_{2k-1}(t) \\
        &~+ A_{2k}(t) G_{2k}(t), \\
        \dot{A}_{2k+1}(t) =&~ -2g_{d}\sqrt{A_{2k+1}(t)A_{2k}(t)} \, \sin \Delta\phi_{2k}(t)  + 2g_{N} \sqrt{A_{2k+1}(t)A_{2k+2}(t)} \, \sin \Delta\phi_{2k+1}(t), \\
        \Delta \dot{\phi}_{2k}(t) =&~ g_{d} \frac{A_{2k+1}(t) - A_{2k}(t)}{\sqrt{A_{2k}(t) A_{2k+1}(t)}} \, \cos \Delta\phi_{2k}(t)  - g_{N}\sqrt{\frac{A_{2k+2}(t)}{A_{2k+1}(t)}} \, \cos \Delta\phi_{2k+1}(t) \\
        &~+ g_{N}\sqrt{\frac{A_{2k-1}(t)}{A_{2k}(t)}} \, \cos \Delta\phi_{2k-1}(t)  + \frac{\alpha}{2} G_{2k}(t), \\
        \Delta \dot{\phi}_{2k+1}(t) =&~ g_{d}\sqrt{\frac{A_{2k}(t)}{A_{2k+1}(t)}} \, \cos \Delta\phi_{2k}(t)  - g_{d}\sqrt{\frac{A_{2k+3}(t)}{A_{2k+2}(t)}} \, \cos \Delta\phi_{2k+2}(t) \\
        &~+ g_{N}\frac{A_{2k+2}(t) - A_{2k+1}(t)}{\sqrt{A_{2k+1}(t)A_{2k+2}(t)}} \, \cos \Delta\phi_{2k+1}(t)  - \frac{\alpha}{2} G_{2k+2}(t), \\
        \dot{n}_{2k}(t) =&~ R - \frac{n_{2k}(t)}{\tau_{s}} - \frac{2 A_{2k}(t) \left[ n_{2k}(t) - 1 \right]}{\tau_{s}}.
    \end{aligned}
\end{align}
The amplitude equations show how intra- and inter-dimer links redistribute intensity around the necklace.
The phase-difference equations show how the same links compete with linewidth-enhancement shifts from the active sites.
The carrier equations close the class-B feedback loop by making the active-site gain depend on the local intensity.
This form isolates the variables that enter the uniform fixed points and the cyclic closure condition.

\section{Fixed points and stability regions} 
\label{sec:Sec3}

A single class-B laser reaches threshold at
\begin{align}
	R_{\mathrm{th}} =&~ \frac{1}{\tau_{s}} \left( \frac{1}{\sigma \tau_{p}} + 1 \right).
\end{align}
The field vanishes for $R < R_{\mathrm{th}}$ and lases for $R > R_{\mathrm{th}}$.
A class-B laser--resonator dimer has equal-amplitude fixed points,
\begin{align}
	A_{1}^{(\mathrm{fp})} =&~ A_{2}^{(\mathrm{fp})} = \frac{1}{2} \left( R \sigma \tau_{p} \tau_{s} - \sigma \tau_{p} - 1 \right),
\end{align}
with carrier inversion $n^{(\mathrm{fp})} = (1 + \sigma \tau_{p}) / (\sigma \tau_{p})$ and phase difference $\Delta\phi^{(\mathrm{fp})} = 0$ or $\pi$~\cite{GuemesFrese2022}.

Our class-B dimer necklace inherits uniform-amplitude fixed points,
\begin{align}
	\begin{aligned}
		A_{2k}^{(\mathrm{fp})} =&~ A_{2k+1}^{(\mathrm{fp})} = \frac{1}{2} \left( R \sigma \tau_{p} \tau_{s} - \sigma \tau_{p} - 1 \right), \\
		n_{2k}^{(\mathrm{fp})} =&~ \frac{1 + \sigma \tau_{p}}{\sigma \tau_{p}}.
	\end{aligned}
\end{align}
These fixed points exist for $R > R_{\mathrm{th}}$ and do not depend on $g_{d}$ or $g_{N}$.
Other, nonuniform fixed points may exist; our classification focuses on these uniform-amplitude configurations.

The phase differences obey the cyclic closure condition $\sum_{k=0}^{2N-1} \Delta\phi_{k} = 2\pi m$, with integer $m$.
Even necklaces support four uniform binary configurations,
\begin{align}
    \begin{aligned}
        \Delta\phi_{k}^{(\mathrm{fp1})} =&~ 0, \\
        \Delta\phi_{k}^{(\mathrm{fp2})} =&~ \pi, \\
        \Delta\phi_{k}^{(\mathrm{fp3})} =&~
        \begin{cases}
            0, & k~\mathrm{even}, \\
            \pi, & k~\mathrm{odd},
        \end{cases} \\
        \Delta\phi_{k}^{(\mathrm{fp4})} =&~
        \begin{cases}
            \pi, & k~\mathrm{even}, \\
            0, & k~\mathrm{odd}.
        \end{cases}
    \end{aligned}
\end{align}
The first configuration synchronizes all sites.
The second imposes a $\pi$ shift between neighboring sites.
The third synchronizes the two sites inside each dimer and imposes a $\pi$ shift between neighboring dimers.
The fourth imposes a $\pi$ shift inside each dimer and synchronizes neighboring dimers.
Odd necklaces exclude the last two alternating configurations, $\Delta\phi_{k}^{(\mathrm{fp3})}$ and $\Delta\phi_{k}^{(\mathrm{fp4})}$, because they accumulate an odd multiple of $\pi$ around the cycle.

We evaluate the Jacobian matrix of the real system at each uniform fixed point and classify local stability from the real parts of its eigenvalues~\cite{Guckenheimer1983,Murray1989,Jordan2007}.
We use $\alpha = 3$, $\sigma = 600~\mathrm{GHz}$, $\tau_{p} = 40~\mathrm{ps}$, and $\tau_{s} = 4~\mathrm{ns}$, and sweep $g_{d}, g_{N} \in [0,250]~\mathrm{ns}^{-1}$~\cite{GuemesFrese2022}.
The four fixed-point stability outcomes define the sixteen regions listed in Table~\ref{tab:Table1}.

\begin{table}[h!]
    \begin{center}
        \begin{tabular}{|c||c c c c|c|} 
            \hline
            \hline
            Region & FP1 & FP2 & FP3 & FP4 & Color \\ \hline\hline
            R0 & Saddle & Saddle & Saddle & Saddle & $\square$ \\ \hline
            R1 & Sink & Saddle & Saddle & Saddle & $\textcolor{cz1}{\mdlgblksquare}$ \\ \hline
            R2 & Saddle & Sink & Saddle & Saddle & $\textcolor{cz2}{\mdlgblksquare}$ \\ \hline
            R3 & Saddle & Saddle & Sink & Saddle & $\textcolor{cz3}{\mdlgblksquare}$ \\ \hline
            R4 & Saddle & Saddle & Saddle & Sink & $\textcolor{cz4}{\mdlgblksquare}$ \\ \hline
            R5 & Sink & Sink & Saddle & Saddle & $\textcolor{cz5}{\mdlgblksquare}$ \\ \hline
            R6 & Sink & Saddle & Sink & Saddle & $\textcolor{cz6}{\mdlgblksquare}$ \\ \hline
            R7 & Sink & Saddle & Saddle & Sink & $\textcolor{cz7}{\mdlgblksquare}$ \\ \hline
            R8 & Saddle & Sink & Sink & Saddle & $\textcolor{cz8}{\mdlgblksquare}$ \\ \hline
            R9 & Saddle & Sink & Saddle & Sink & $\textcolor{cz9}{\mdlgblksquare}$ \\ \hline
            R10 & Saddle & Saddle & Sink & Sink & $\textcolor{cz10}{\mdlgblksquare}$ \\ \hline
            R11 & Sink & Sink & Sink & Saddle & $\textcolor{cz11}{\mdlgblksquare}$ \\ \hline
            R12 & Sink & Sink & Saddle & Sink & $\textcolor{cz12}{\mdlgblksquare}$ \\ \hline
            R13 & Sink & Saddle & Sink & Sink & $\textcolor{cz13}{\mdlgblksquare}$ \\ \hline
            R14 & Saddle & Sink & Sink & Sink & $\textcolor{cz14}{\mdlgblksquare}$ \\ \hline
            R15 & Sink & Sink & Sink & Sink & $\textcolor{cz15}{\mdlgblksquare}$ \\ \hline
            \hline
        \end{tabular}
        \caption{Stability regions defined by the four uniform fixed-point configurations. 
        Region R0 has no stable uniform fixed point among the tested configurations.}
        \label{tab:Table1}
    \end{center}
\end{table}

Region R0 contains no stable uniform fixed point among FP1--FP4.
Regions R1--R4 stabilize one configuration, R5--R10 stabilize two, R11--R14 stabilize three, and R15 stabilizes all four.
The phase-closure constraint determines which of these regions can appear for each necklace parity.

\begin{figure*}
    \centering
    \includegraphics[width=\textwidth]{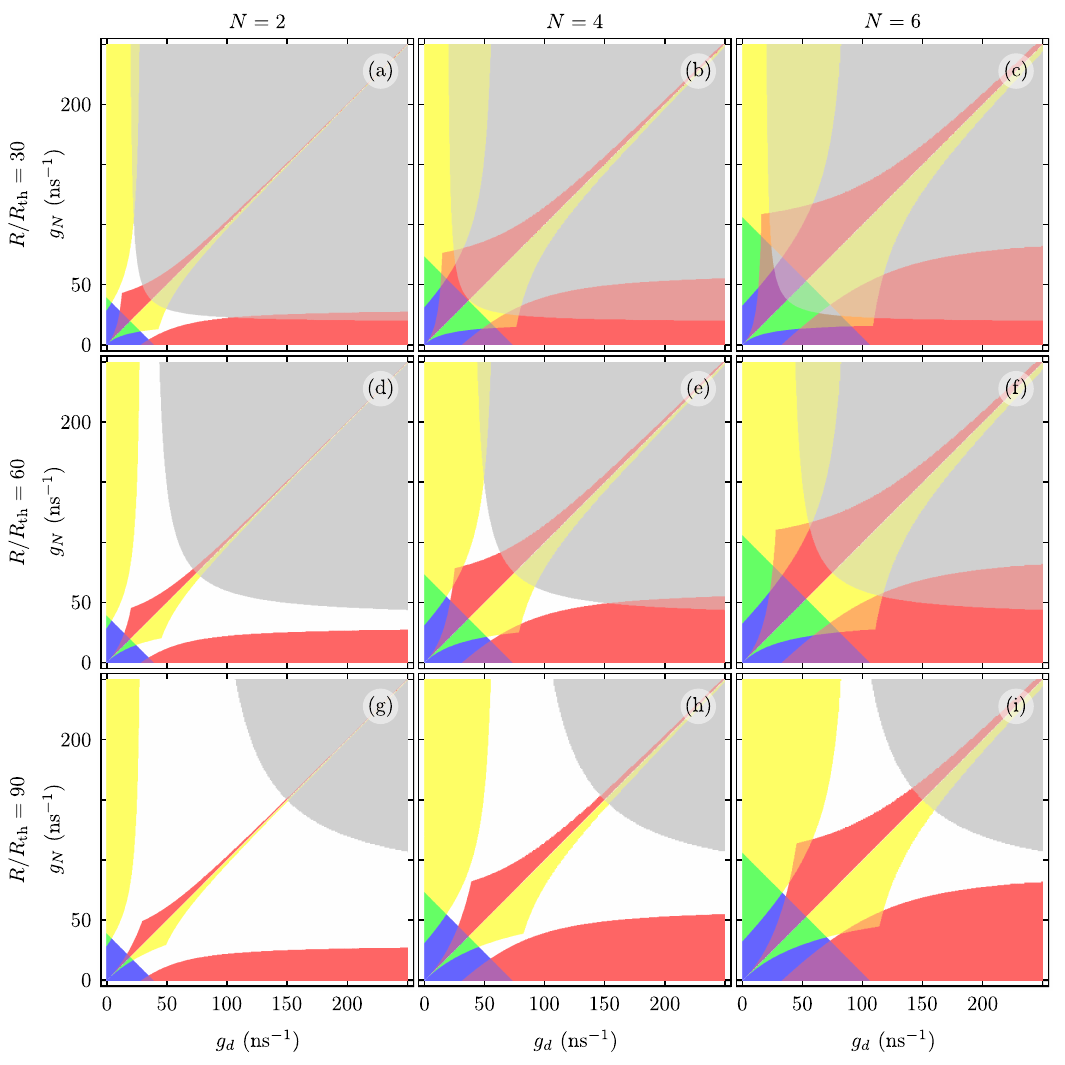}
    \caption{Fixed-point stability map for even necklaces with $N = 2,4,6$ at $R / R_{\mathrm{th}} = 30,60,90$. Colors follow Table~\ref{tab:Table1}.}
    \label{fig:Fig2}
\end{figure*}

Figure~\ref{fig:Fig2} shows the stability regions for even necklaces with $N = 2$, $4$, and $6$.
Even necklaces can realize all regions except R5 in the explored parameter range.
For fixed $N$, increasing $R/R_{\mathrm{th}}$ expands stable domains and reduces R0.
For fixed pump ratio, increasing $N$ shifts regions R2, R8, and R9 toward larger couplings, while regions R1, R3, and R4 keep approximately the same area.

\begin{figure*}
    \centering
    \includegraphics[width=\textwidth]{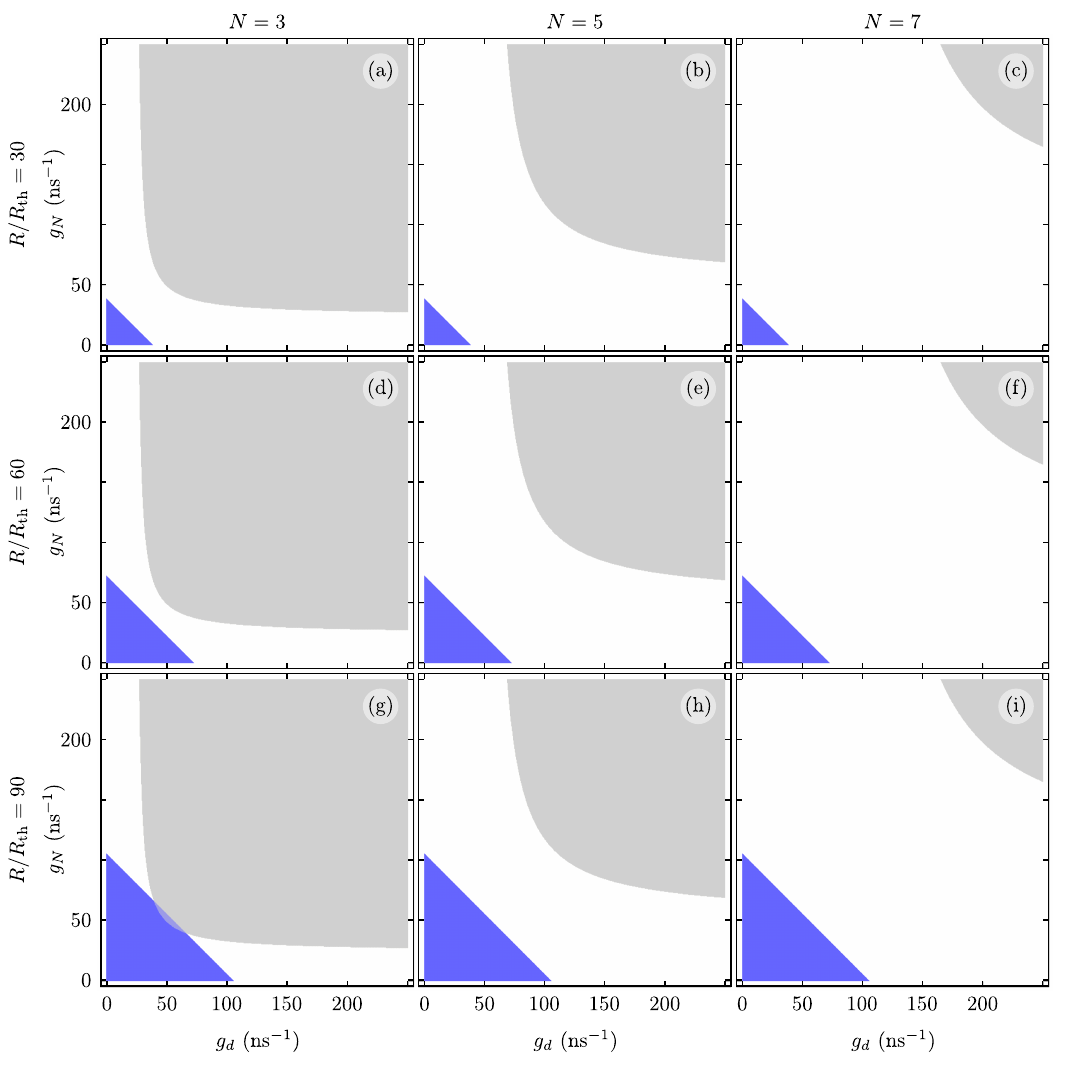}
    \caption{Fixed-point stability map for odd necklaces with $N = 3,5,7$ at $R / R_{\mathrm{th}} = 30,60,90$. Colors follow Table~\ref{tab:Table1}.}
    \label{fig:Fig3}
\end{figure*}

Figure~\ref{fig:Fig3} shows the stability regions for odd necklaces with $N = 3$, $5$, and $7$.
Odd necklaces realize only R0, R1, R2, and R5 because FP3 and FP4 violate cyclic phase closure.
For fixed $N$, increasing $R/R_{\mathrm{th}}$ suppresses R0 and expands the stable regions.

We complement the local stability maps with long-time propagation along selected lines in the $(g_{d},g_{N})$ plane.
For each parameter pair, we propagate randomly-generated initial conditions.
The integration time is $95~\mathrm{ns}$.
Single-attractor regions collapse to one asymptotic amplitude--phase value, multistable regions split into distinct branches, and chaotic or nonstationary regions produce dispersed final values.
Transitions between these dynamical behaviors occur near the region boundaries in the $(g_{d},g_{N},R/R_{\mathrm{th}})$ stability volume.

\subsection{Two-dimer Necklace}

\begin{figure*}
    \centering
    \includegraphics[width=\textwidth]{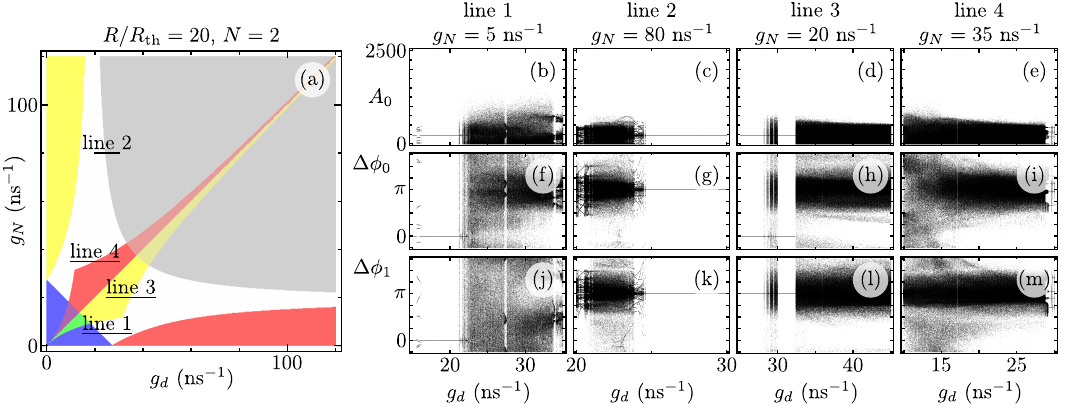}
    \caption{Two-dimer necklace at $R/R_{\mathrm{th}} = 20$. 
        (a) Fixed-point stability map in the $(g_{d},g_{N})$ plane. 
        (b)--(e) Long-time propagation along lines L1--L4.}
    \label{fig:Fig4}
\end{figure*}

\begin{figure*}
    \centering
    \includegraphics[width=\textwidth]{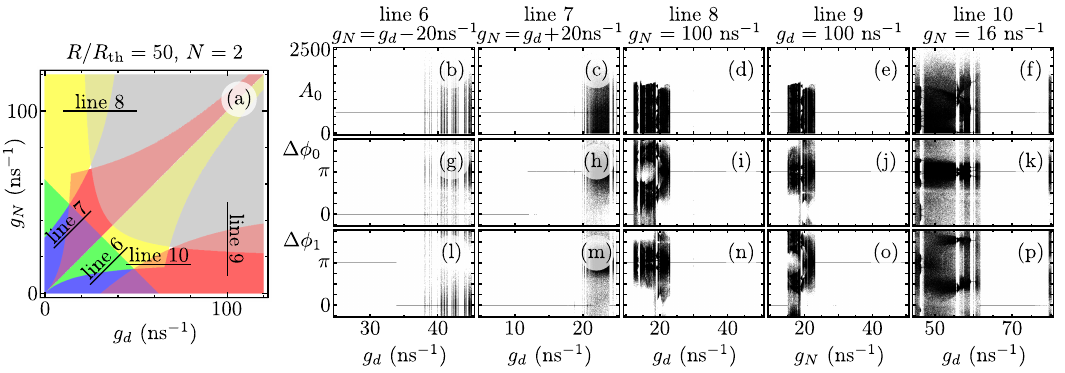}
    \caption{Two-dimer necklace at $R/R_{\mathrm{th}} = 50$. 
        (a) Fixed-point stability map in the $(g_{d},g_{N})$ plane. 
        (b)--(f) Long-time propagation along lines L6--L10.}
    \label{fig:Fig5}
\end{figure*}

\begin{figure*}
    \centering
    \includegraphics[width=\textwidth]{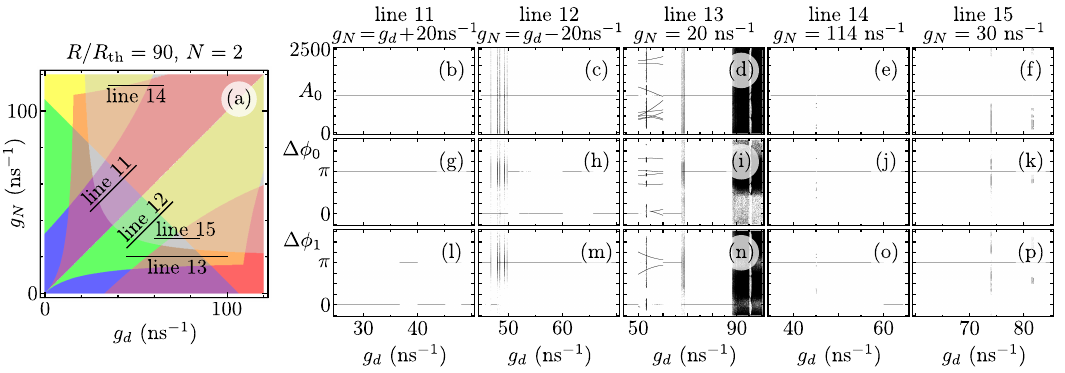}
    \caption{Two-dimer necklace at $R/R_{\mathrm{th}} = 90$. 
        (a) Fixed-point stability map in the $(g_{d},g_{N})$ plane. 
        (b)--(f) Long-time propagation along lines L11--L15.}
    \label{fig:Fig6}
\end{figure*}

The two-dimer necklace gives the smallest even cyclic array and supports the four phase configurations FP1--FP4.
Figures~\ref{fig:Fig4}(a)--\ref{fig:Fig6}(a) show the stability regions in the $(g_{d},g_{N})$ coupling plane.
The remaining panels show the asymptotic values of $A_{0}$, $\Delta\phi_{0}$, and $\Delta\phi_{1}$ for randomly-generated initial conditions along the indicated lines.

At $R/R_{\mathrm{th}} = 20$, region R0 occupies a broad part of the coupling plane, Fig.~\ref{fig:Fig4}(a).
All four sweeps L1--L4 intersect R0.
Sweeps L1 and L3 cross from a stable region into R0, while sweeps L2 and L4 cross from R0 into a stable region.
Inside R0, the propagated trajectories show dispersed amplitude and phase-difference values, consistent with irregular long-time dynamics.
Sweeps L1 and L3 also show intervals where the dispersed dynamics collapse onto stable branches before spreading again inside R0.
This collapse is pronounced along L3, while along L1 it may indicate multistability, a long transient, or a narrow regular window inside the irregular region.
Sweeps L2 and L4 show a more abrupt transition between dispersed dynamics in R0 and stable branches outside R0.

At $R/R_{\mathrm{th}} = 50$, R0 shrinks and stable regions dominate the coupling plane, Fig.~\ref{fig:Fig5}(a).
Sweeps L6--L10 show abrupt transitions between stable and dispersed asymptotic values as the lines cross region boundaries.
Several sweeps show signs of multistability through the coexistence of multiple amplitude and phase-difference branches.
This behavior is clearest along L10, where the line crosses boundaries between regions with different numbers of stable fixed-point configurations.
The propagated trajectories then split into initial-condition-dependent branches.

At $R/R_{\mathrm{th}} = 90$, R0 disappears from the explored coupling plane, Fig.~\ref{fig:Fig6}(a).
All visible regions contain at least one stable uniform fixed point, so the broad dispersed intervals observed at lower pump ratio disappear.
Sweeps L11--L14 mainly switch between stable asymptotic branches associated with different stable phase configurations.
Sweep L15 shows the clearest multistability signal at this pump ratio.
It crosses a boundary where the system changes from two to three stable fixed-point configurations, and the propagated trajectories split into coexisting amplitude and phase branches.

For $N = 2$, all sweeps show abrupt phase-configuration changes at stability-region boundaries.
The pump ratio controls whether those changes occur through broad irregular intervals or between concentrated asymptotic branches.
Low pump ratio leaves a broad R0 region where irregular dynamics and narrow regular intervals coexist.
Larger pump ratios suppress R0 and favor stable switching between phase configurations.
Multistability appears most clearly where the number of stable fixed-point configurations changes.
	
\subsection{Three-dimer Necklace}

\begin{figure}
    \centering
    \includegraphics[scale=1]{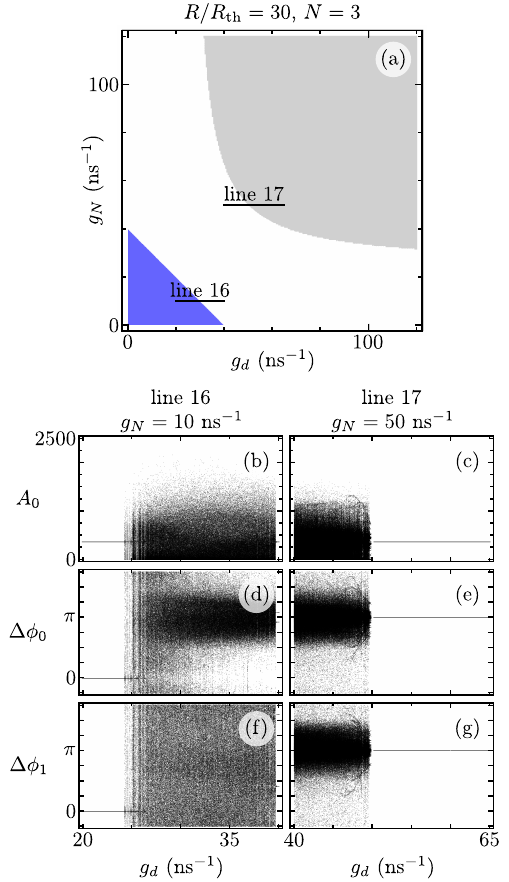}
    \caption{Three-dimer necklace at $R/R_{\mathrm{th}} = 30$. 
        (a) Fixed-point stability map in the $(g_{d},g_{N})$ plane with regions R0, R1, and R2. 
        (b)--(g) Long-time propagation along lines L16 and L17.}
    \label{fig:Fig7}
\end{figure}

\begin{figure}
    \centering
    \includegraphics[scale=1]{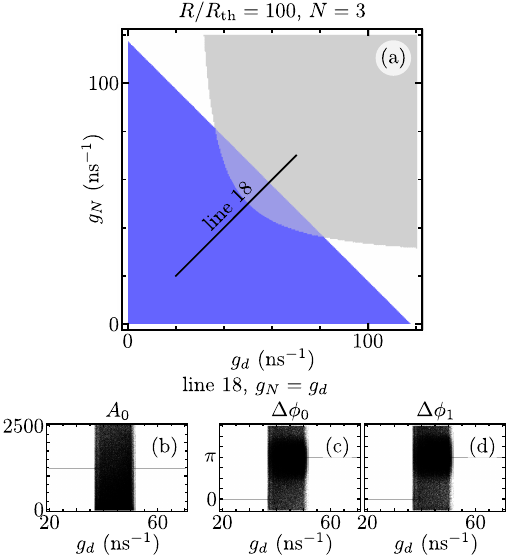}
    \caption{Three-dimer necklace at $R/R_{\mathrm{th}} = 100$. 
        (a) Fixed-point stability map in the $(g_{d},g_{N})$ plane with regions R0, R1, R2, and R5. 
        (b)--(d) Long-time propagation along line L18.}
    \label{fig:Fig8}
\end{figure}

The three-dimer necklace gives the smallest odd cyclic array and excludes the alternating configurations FP3 and FP4.
Only FP1 and FP2 satisfy cyclic phase closure, so the stability maps can contain only regions R0, R1, R2, and R5.
Figure~\ref{fig:Fig7}(a) shows regions R0, R1, and R2 at $R/R_{\mathrm{th}} = 30$, while Fig.~\ref{fig:Fig8}(a) shows regions R0, R1, R2, and R5 at $R/R_{\mathrm{th}} = 100$.
The remaining panels show the asymptotic values of $A_{0}$, $\Delta\phi_{0}$, and $\Delta\phi_{1}$ for randomly-generated initial conditions along the indicated lines.

At $R/R_{\mathrm{th}} = 30$, Fig.~\ref{fig:Fig7}(a) shows only regions R0, R1, and R2.
Line L16 crosses from R1 into R0, showing the loss of stability of FP1, and L17 crosses from R0 into R2, showing stabilization into FP2.
Both sweeps show abrupt changes near the region boundaries, while the trajectories inside R0 disperse and show irregular dynamics.

At $R/R_{\mathrm{th}} = 100$, Fig.~\ref{fig:Fig8}(a) shows regions R0, R1, R2, and R5.
Region R0 shrinks, and region R5 appears between R1 and R2.
Line L18 follows the diagonal line $g_{N} = g_{d}$ and crosses the R1--R5--R2 sequence.
The trajectories show abrupt changes of amplitude and phase difference near the R1--R5 boundary.
The spread of final values near the R5--R2 transition suggests a multistable crossover.

For $N = 3$, cyclic phase closure restricts the observed dynamics to the fully synchronized and uniformly $\pi$-shifted fixed-point configurations.
Low pump ratio leaves a broad R0 region where the trajectories disperse and show irregular dynamics.
Larger pump ratio suppresses R0 and opens region R5, where both allowed phase configurations are stable.
The spread of final values near the R5--R2 transition suggests that multistability emerges when both stable fixed-point configurations become available.

\section{Conclusion} 
\label{sec:Sec4}

We constructed a class-B dimer necklace by placing carrier-inversion dynamics on the active sites of a cyclic $\mathcal{PT}$-symmetric dimer necklace.
The resulting amplitude--phase--carrier model combines intra-dimer coupling, inter-dimer coupling, linewidth-enhancement phase shifts, and local carrier feedback in the same dynamical system.
Our construction extends the linear necklace from symmetry-protected mode selection to nonlinear carrier-mediated stability and irregular lasing.

Uniform fixed points inherit the class-B laser--resonator amplitude and carrier inversion, while cyclic closure restricts their phase configurations.
Even necklaces support full synchronization, a uniform $\pi$ shift, and two alternating configurations.
Odd necklaces exclude the alternating configurations and retain only the fully synchronized and uniformly $\pi$-shifted configurations.
Linear stability of these fixed points partitions the coupling plane into sixteen possible regions, with R0 marking the absence of stable uniform fixed points among the tested configurations.

Long-time propagation shows how these stability regions organize the observed dynamics.
For the two-dimer necklace, low pump ratio leaves a broad R0 region where irregular dynamics appear, while larger pump ratios suppress R0 and favor stable switching between phase configurations.
Multistability appears most clearly near boundaries where the number of stable fixed-point configurations changes.
For the three-dimer necklace, cyclic phase closure restricts the dynamics to the two allowed phase configurations, and larger pump ratio opens region R5, where both configurations are stable.
The spread of final values near the R5--R2 transition suggests a multistable crossover.

These results identify cyclic phase closure as the constraint that separates even and odd routes to carrier-mediated nonlinear dynamics in class-B dimer necklaces.
The pump ratio controls the size of the no-stable-fixed-point region, while the intra- and inter-dimer couplings select the accessible phase configurations and their stability.
The class-B dimer necklace therefore provides a minimal setting where coupling topology and carrier feedback organize stable lasing, multistability, and irregular lasing in integrated photonic arrays.


\section*{Funding}
This work received no funding.

\section*{Acknowledgment}
B.~M.~R.~L. acknowledges support and hospitality as an affiliate visiting colleague at the Department of Physics and Astronomy, University of New Mexico.

\section*{Disclosures} 
The authors declare no conflicts of interest.

\section*{Data availability} 
The numerical data underlying the figures were generated from the model equations and parameter values stated in the text and are available from the corresponding author upon reasonable request. 
No external datasets were used or analyzed in this work.



\begin{thebibliography}{42}%
\makeatletter
\providecommand \@ifxundefined [1]{%
 \@ifx{#1\undefined}
}%
\providecommand \@ifnum [1]{%
 \ifnum #1\expandafter \@firstoftwo
 \else \expandafter \@secondoftwo
 \fi
}%
\providecommand \@ifx [1]{%
 \ifx #1\expandafter \@firstoftwo
 \else \expandafter \@secondoftwo
 \fi
}%
\providecommand \natexlab [1]{#1}%
\providecommand \enquote  [1]{``#1''}%
\providecommand \bibnamefont  [1]{#1}%
\providecommand \bibfnamefont [1]{#1}%
\providecommand \citenamefont [1]{#1}%
\providecommand \href@noop [0]{\@secondoftwo}%
\providecommand \href [0]{\begingroup \@sanitize@url \@href}%
\providecommand \@href[1]{\@@startlink{#1}\@@href}%
\providecommand \@@href[1]{\endgroup#1\@@endlink}%
\providecommand \@sanitize@url [0]{\catcode `\\12\catcode `\$12\catcode
  `\&12\catcode `\#12\catcode `\^12\catcode `\_12\catcode `\%12\relax}%
\providecommand \@@startlink[1]{}%
\providecommand \@@endlink[0]{}%
\providecommand \url  [0]{\begingroup\@sanitize@url \@url }%
\providecommand \@url [1]{\endgroup\@href {#1}{\urlprefix }}%
\providecommand \urlprefix  [0]{URL }%
\providecommand \Eprint [0]{\href }%
\providecommand \doibase [0]{https://doi.org/}%
\providecommand \selectlanguage [0]{\@gobble}%
\providecommand \bibinfo  [0]{\@secondoftwo}%
\providecommand \bibfield  [0]{\@secondoftwo}%
\providecommand \translation [1]{[#1]}%
\providecommand \BibitemOpen [0]{}%
\providecommand \bibitemStop [0]{}%
\providecommand \bibitemNoStop [0]{.\EOS\space}%
\providecommand \EOS [0]{\spacefactor3000\relax}%
\providecommand \BibitemShut  [1]{\csname bibitem#1\endcsname}%
\let\auto@bib@innerbib\@empty
\bibitem [{\citenamefont {Brunner}\ \emph {et~al.}(2020)\citenamefont
  {Brunner}, \citenamefont {Marandi}, \citenamefont {Bogaerts},\ and\
  \citenamefont {Ozcan}}]{Brunner2020}%
  \BibitemOpen
  \bibfield  {author} {\bibinfo {author} {\bibfnamefont {D.}~\bibnamefont
  {Brunner}}, \bibinfo {author} {\bibfnamefont {A.}~\bibnamefont {Marandi}},
  \bibinfo {author} {\bibfnamefont {W.}~\bibnamefont {Bogaerts}},\ and\
  \bibinfo {author} {\bibfnamefont {A.}~\bibnamefont {Ozcan}},\ }\bibfield
  {title} {\bibinfo {title} {Photonics for computing and computing for
  photonics},\ }\href {https://doi.org/10.1515/nanoph-2020-0470} {\bibfield
  {journal} {\bibinfo  {journal} {Nanophotonics}\ }\textbf {\bibinfo {volume}
  {9}},\ \bibinfo {pages} {4053} (\bibinfo {year} {2020})}\BibitemShut
  {NoStop}%
\bibitem [{\citenamefont {Pelucchi}\ \emph {et~al.}(2022)\citenamefont
  {Pelucchi}, \citenamefont {Aharonovich}, \citenamefont {Englund},
  \citenamefont {Figueroa}, \citenamefont {Gong}, \citenamefont {Hannes},
  \citenamefont {Liu}, \citenamefont {Lu}, \citenamefont {Matsuda},
  \citenamefont {Pan}, \citenamefont {Schreck}, \citenamefont {Sciarrino},
  \citenamefont {Silberhorn}, \citenamefont {Wang},\ and\ \citenamefont
  {J{\"o}ns}}]{Pelucchi2022}%
  \BibitemOpen
  \bibfield  {author} {\bibinfo {author} {\bibfnamefont {E.}~\bibnamefont
  {Pelucchi}}, \bibinfo {author} {\bibfnamefont {G.~F.~I.}\ \bibnamefont
  {Aharonovich}}, \bibinfo {author} {\bibfnamefont {D.}~\bibnamefont
  {Englund}}, \bibinfo {author} {\bibfnamefont {E.}~\bibnamefont {Figueroa}},
  \bibinfo {author} {\bibfnamefont {Q.}~\bibnamefont {Gong}}, \bibinfo {author}
  {\bibfnamefont {H.}~\bibnamefont {Hannes}}, \bibinfo {author} {\bibfnamefont
  {J.}~\bibnamefont {Liu}}, \bibinfo {author} {\bibfnamefont {C.-Y.}\
  \bibnamefont {Lu}}, \bibinfo {author} {\bibfnamefont {N.}~\bibnamefont
  {Matsuda}}, \bibinfo {author} {\bibfnamefont {J.-W.}\ \bibnamefont {Pan}},
  \bibinfo {author} {\bibfnamefont {F.}~\bibnamefont {Schreck}}, \bibinfo
  {author} {\bibfnamefont {F.}~\bibnamefont {Sciarrino}}, \bibinfo {author}
  {\bibfnamefont {C.}~\bibnamefont {Silberhorn}}, \bibinfo {author}
  {\bibfnamefont {J.}~\bibnamefont {Wang}},\ and\ \bibinfo {author}
  {\bibfnamefont {K.~D.}\ \bibnamefont {J{\"o}ns}},\ }\bibfield  {title}
  {\bibinfo {title} {The potential and global outlook of integrated photonics
  for quantum technologies},\ }\href
  {https://doi.org/10.1038/s42254-021-00398-z} {\bibfield  {journal} {\bibinfo
  {journal} {Nat. Rev. Phys.}\ }\textbf {\bibinfo {volume} {4}},\ \bibinfo
  {pages} {194} (\bibinfo {year} {2022})}\BibitemShut {NoStop}%
\bibitem [{\citenamefont {{PsiQuantum Team}}(2025)}]{PsiQuantum2025}%
  \BibitemOpen
  \bibfield  {author} {\bibinfo {author} {\bibnamefont {{PsiQuantum Team}}},\
  }\bibfield  {title} {\bibinfo {title} {A manufacturable platform for photonic
  quantum computing},\ }\href {https://doi.org/10.1038/s41586-025-08820-7}
  {\bibfield  {journal} {\bibinfo  {journal} {Nature}\ }\textbf {\bibinfo
  {volume} {641}},\ \bibinfo {pages} {876} (\bibinfo {year}
  {2025})}\BibitemShut {NoStop}%
\bibitem [{\citenamefont {Chovan}\ and\ \citenamefont
  {Uherek}(2018)}]{Chovan2018}%
  \BibitemOpen
  \bibfield  {author} {\bibinfo {author} {\bibfnamefont {J.}~\bibnamefont
  {Chovan}}\ and\ \bibinfo {author} {\bibfnamefont {F.}~\bibnamefont
  {Uherek}},\ }\bibfield  {title} {\bibinfo {title} {Photonic integrated
  circuits for communication systems},\ }\href
  {https://doi.org/10.13164/re.2018.0357} {\bibfield  {journal} {\bibinfo
  {journal} {Radioengineering}\ }\textbf {\bibinfo {volume} {27}},\ \bibinfo
  {pages} {357} (\bibinfo {year} {2018})}\BibitemShut {NoStop}%
\bibitem [{\citenamefont {Glick}\ \emph {et~al.}(2023)\citenamefont {Glick},
  \citenamefont {Liao},\ and\ \citenamefont {Schmidtke}}]{Glick2023}%
  \BibitemOpen
  \bibinfo {editor} {\bibfnamefont {M.}~\bibnamefont {Glick}}, \bibinfo
  {editor} {\bibfnamefont {L.}~\bibnamefont {Liao}},\ and\ \bibinfo {editor}
  {\bibfnamefont {K.}~\bibnamefont {Schmidtke}},\ eds.,\ \href
  {https://doi.org/10.1016/C2020-0-03648-0} {\emph {\bibinfo {title}
  {Integrated Photonics for Data Communication Applications}}}\ (\bibinfo
  {publisher} {Elsevier},\ \bibinfo {year} {2023})\BibitemShut {NoStop}%
\bibitem [{\citenamefont {Arafin}\ and\ \citenamefont
  {Coldren}(2018)}]{Arafin2018}%
  \BibitemOpen
  \bibfield  {author} {\bibinfo {author} {\bibfnamefont {S.}~\bibnamefont
  {Arafin}}\ and\ \bibinfo {author} {\bibfnamefont {L.~A.}\ \bibnamefont
  {Coldren}},\ }\bibfield  {title} {\bibinfo {title} {Advanced {InP} photonic
  integrated circuits for communication and sensing},\ }\href
  {https://doi.org/10.1109/JSTQE.2017.2754583} {\bibfield  {journal} {\bibinfo
  {journal} {J. Sel. Top. Quant.}\ }\textbf {\bibinfo {volume} {24}},\ \bibinfo
  {pages} {1} (\bibinfo {year} {2018})}\BibitemShut {NoStop}%
\bibitem [{\citenamefont {Zhou}\ \emph {et~al.}(2023)\citenamefont {Zhou},
  \citenamefont {Ou}, \citenamefont {Fang}, \citenamefont {Alkhazraji},
  \citenamefont {Xu}, \citenamefont {Wan},\ and\ \citenamefont
  {Bowers}}]{Zhou2023}%
  \BibitemOpen
  \bibfield  {author} {\bibinfo {author} {\bibfnamefont {Z.}~\bibnamefont
  {Zhou}}, \bibinfo {author} {\bibfnamefont {X.}~\bibnamefont {Ou}}, \bibinfo
  {author} {\bibfnamefont {Y.}~\bibnamefont {Fang}}, \bibinfo {author}
  {\bibfnamefont {E.}~\bibnamefont {Alkhazraji}}, \bibinfo {author}
  {\bibfnamefont {R.}~\bibnamefont {Xu}}, \bibinfo {author} {\bibfnamefont
  {Y.}~\bibnamefont {Wan}},\ and\ \bibinfo {author} {\bibfnamefont {J.~E.}\
  \bibnamefont {Bowers}},\ }\bibfield  {title} {\bibinfo {title} {Prospects and
  applications of on-chip lasers},\ }\href
  {https://doi.org/10.1186/s43593-022-00027-x} {\bibfield  {journal} {\bibinfo
  {journal} {eLight}\ }\textbf {\bibinfo {volume} {3}},\ \bibinfo {pages} {1}
  (\bibinfo {year} {2023})}\BibitemShut {NoStop}%
\bibitem [{\citenamefont {Gourley}\ \emph {et~al.}(1991)\citenamefont
  {Gourley}, \citenamefont {Warren}, \citenamefont {Hadley}, \citenamefont
  {Vawter}, \citenamefont {Brennan},\ and\ \citenamefont
  {Hammons}}]{Gourley1991}%
  \BibitemOpen
  \bibfield  {author} {\bibinfo {author} {\bibfnamefont {P.~L.}\ \bibnamefont
  {Gourley}}, \bibinfo {author} {\bibfnamefont {M.~E.}\ \bibnamefont {Warren}},
  \bibinfo {author} {\bibfnamefont {G.~R.}\ \bibnamefont {Hadley}}, \bibinfo
  {author} {\bibfnamefont {G.~A.}\ \bibnamefont {Vawter}}, \bibinfo {author}
  {\bibfnamefont {T.~M.}\ \bibnamefont {Brennan}},\ and\ \bibinfo {author}
  {\bibfnamefont {B.~E.}\ \bibnamefont {Hammons}},\ }\bibfield  {title}
  {\bibinfo {title} {Coherent beams from high efficiency two‐dimensional
  surface‐emitting semiconductor laser arrays},\ }\href
  {https://doi.org/10.1063/1.104468} {\bibfield  {journal} {\bibinfo  {journal}
  {Appl. Phys. Lett.}\ }\textbf {\bibinfo {volume} {58}},\ \bibinfo {pages}
  {890} (\bibinfo {year} {1991})}\BibitemShut {NoStop}%
\bibitem [{\citenamefont {Garc\'ia-{O}jalvo}\ \emph {et~al.}(1999)\citenamefont
  {Garc\'ia-{O}jalvo}, \citenamefont {Casademont}, \citenamefont {Torrent},
  \citenamefont {Mirasso},\ and\ \citenamefont {Sancho}}]{Garcia1999}%
  \BibitemOpen
  \bibfield  {author} {\bibinfo {author} {\bibfnamefont {J.}~\bibnamefont
  {Garc\'ia-{O}jalvo}}, \bibinfo {author} {\bibfnamefont {J.}~\bibnamefont
  {Casademont}}, \bibinfo {author} {\bibfnamefont {M.~C.}\ \bibnamefont
  {Torrent}}, \bibinfo {author} {\bibfnamefont {C.~R.}\ \bibnamefont
  {Mirasso}},\ and\ \bibinfo {author} {\bibfnamefont {J.~M.}\ \bibnamefont
  {Sancho}},\ }\bibfield  {title} {\bibinfo {title} {Coherence and
  synchronization in diode-laser arrays with delayed global coupling},\ }\href
  {https://doi.org/10.1142/S021812749900167X} {\bibfield  {journal} {\bibinfo
  {journal} {Int. J. Bifurcat. Chaos}\ }\textbf {\bibinfo {volume} {09}},\
  \bibinfo {pages} {2225} (\bibinfo {year} {1999})}\BibitemShut {NoStop}%
\bibitem [{\citenamefont {Winful}\ and\ \citenamefont
  {Rahman}(1990)}]{Winful1990}%
  \BibitemOpen
  \bibfield  {author} {\bibinfo {author} {\bibfnamefont {H.~G.}\ \bibnamefont
  {Winful}}\ and\ \bibinfo {author} {\bibfnamefont {L.}~\bibnamefont
  {Rahman}},\ }\bibfield  {title} {\bibinfo {title} {Synchronized chaos and
  spatiotemporal chaos in arrays of coupled lasers},\ }\href
  {https://doi.org/10.1103/PhysRevLett.65.1575} {\bibfield  {journal} {\bibinfo
   {journal} {Phys. Rev. Lett.}\ }\textbf {\bibinfo {volume} {65}},\ \bibinfo
  {pages} {1575} (\bibinfo {year} {1990})}\BibitemShut {NoStop}%
\bibitem [{\citenamefont {Winful}(1992)}]{Winful1992}%
  \BibitemOpen
  \bibfield  {author} {\bibinfo {author} {\bibfnamefont {H.~G.}\ \bibnamefont
  {Winful}},\ }\bibfield  {title} {\bibinfo {title} {Instability threshold for
  an array of coupled semiconductor lasers},\ }\href
  {https://doi.org/10.1103/PhysRevA.46.6093} {\bibfield  {journal} {\bibinfo
  {journal} {Phys. Rev. A}\ }\textbf {\bibinfo {volume} {46}},\ \bibinfo
  {pages} {6093} (\bibinfo {year} {1992})}\BibitemShut {NoStop}%
\bibitem [{\citenamefont {Thornburg}\ \emph {et~al.}(1997)\citenamefont
  {Thornburg}, \citenamefont {M\"oller}, \citenamefont {Roy}, \citenamefont
  {Carr}, \citenamefont {Li},\ and\ \citenamefont {Erneux}}]{Thornburg1997}%
  \BibitemOpen
  \bibfield  {author} {\bibinfo {author} {\bibfnamefont {K.~S.}\ \bibnamefont
  {Thornburg}}, \bibinfo {author} {\bibfnamefont {M.}~\bibnamefont {M\"oller}},
  \bibinfo {author} {\bibfnamefont {R.}~\bibnamefont {Roy}}, \bibinfo {author}
  {\bibfnamefont {T.~W.}\ \bibnamefont {Carr}}, \bibinfo {author}
  {\bibfnamefont {R.-D.}\ \bibnamefont {Li}},\ and\ \bibinfo {author}
  {\bibfnamefont {T.}~\bibnamefont {Erneux}},\ }\bibfield  {title} {\bibinfo
  {title} {Chaos and coherence in coupled lasers},\ }\href
  {https://doi.org/10.1103/PhysRevE.55.3865} {\bibfield  {journal} {\bibinfo
  {journal} {Phys. Rev. E}\ }\textbf {\bibinfo {volume} {55}},\ \bibinfo
  {pages} {3865} (\bibinfo {year} {1997})}\BibitemShut {NoStop}%
\bibitem [{\citenamefont {Arroyo-Almanza}\ \emph {et~al.}(2012)\citenamefont
  {Arroyo-Almanza}, \citenamefont {Pisarchik},\ and\ \citenamefont
  {Ruiz-{O}liveras}}]{Arroyo2012}%
  \BibitemOpen
  \bibfield  {author} {\bibinfo {author} {\bibfnamefont {D.~A.}\ \bibnamefont
  {Arroyo-Almanza}}, \bibinfo {author} {\bibfnamefont {A.~N.}\ \bibnamefont
  {Pisarchik}},\ and\ \bibinfo {author} {\bibfnamefont {F.~R.}\ \bibnamefont
  {Ruiz-{O}liveras}},\ }\bibfield  {title} {\bibinfo {title} {Route to chaos in
  a ring of three unidirectionally-coupled semiconductor lasers},\ }\href
  {https://doi.org/10.1109/LPT.2012.2184746} {\bibfield  {journal} {\bibinfo
  {journal} {IEEE Photonic. Tech. Let.}\ }\textbf {\bibinfo {volume} {24}},\
  \bibinfo {pages} {681} (\bibinfo {year} {2012})}\BibitemShut {NoStop}%
\bibitem [{\citenamefont {Adams}\ \emph {et~al.}(2022)\citenamefont {Adams},
  \citenamefont {Seyab}, \citenamefont {Henning}, \citenamefont {Susanto},\
  and\ \citenamefont {Vaughan}}]{Adams2022}%
  \BibitemOpen
  \bibfield  {author} {\bibinfo {author} {\bibfnamefont {M.}~\bibnamefont
  {Adams}}, \bibinfo {author} {\bibfnamefont {R.~A.}\ \bibnamefont {Seyab}},
  \bibinfo {author} {\bibfnamefont {I.}~\bibnamefont {Henning}}, \bibinfo
  {author} {\bibfnamefont {H.}~\bibnamefont {Susanto}},\ and\ \bibinfo {author}
  {\bibfnamefont {M.}~\bibnamefont {Vaughan}},\ }\bibfield  {title} {\bibinfo
  {title} {Dynamics of evanescently-coupled laser pairs with unequal pumping:
  Analysis using a three-variable reduction of the coupled rate equations},\
  }\href {https://doi.org/10.1109/JSTQE.2021.3076083} {\bibfield  {journal}
  {\bibinfo  {journal} {{IEEE} J. Sel. Top. Quant.}\ }\textbf {\bibinfo
  {volume} {28}},\ \bibinfo {pages} {1} (\bibinfo {year} {2022})}\BibitemShut
  {NoStop}%
\bibitem [{\citenamefont {El-Ganainy}\ \emph {et~al.}(2018)\citenamefont
  {El-Ganainy}, \citenamefont {Makris}, \citenamefont {Khajavikhan},
  \citenamefont {Musslimani}, \citenamefont {Rotter},\ and\ \citenamefont
  {Christodoulides}}]{ElGanainy2018}%
  \BibitemOpen
  \bibfield  {author} {\bibinfo {author} {\bibfnamefont {R.}~\bibnamefont
  {El-Ganainy}}, \bibinfo {author} {\bibfnamefont {K.~G.}\ \bibnamefont
  {Makris}}, \bibinfo {author} {\bibfnamefont {M.}~\bibnamefont {Khajavikhan}},
  \bibinfo {author} {\bibfnamefont {Z.~H.}\ \bibnamefont {Musslimani}},
  \bibinfo {author} {\bibfnamefont {S.}~\bibnamefont {Rotter}},\ and\ \bibinfo
  {author} {\bibfnamefont {D.~N.}\ \bibnamefont {Christodoulides}},\ }\bibfield
   {title} {\bibinfo {title} {Non-{H}ermitian physics and {PT} symmetry},\
  }\href {https://doi.org/10.1038/nphys4323} {\bibfield  {journal} {\bibinfo
  {journal} {Nat. Phys.}\ }\textbf {\bibinfo {volume} {14}},\ \bibinfo {pages}
  {11} (\bibinfo {year} {2018})}\BibitemShut {NoStop}%
\bibitem [{\citenamefont {Bandres}\ \emph {et~al.}(2018)\citenamefont
  {Bandres}, \citenamefont {Wittek}, \citenamefont {Harari}, \citenamefont
  {Parto}, \citenamefont {Ren}, \citenamefont {Segev}, \citenamefont
  {Christodoulides},\ and\ \citenamefont {Khajavikhan}}]{Bandres2018}%
  \BibitemOpen
  \bibfield  {author} {\bibinfo {author} {\bibfnamefont {M.~A.}\ \bibnamefont
  {Bandres}}, \bibinfo {author} {\bibfnamefont {S.}~\bibnamefont {Wittek}},
  \bibinfo {author} {\bibfnamefont {G.}~\bibnamefont {Harari}}, \bibinfo
  {author} {\bibfnamefont {M.}~\bibnamefont {Parto}}, \bibinfo {author}
  {\bibfnamefont {J.}~\bibnamefont {Ren}}, \bibinfo {author} {\bibfnamefont
  {M.}~\bibnamefont {Segev}}, \bibinfo {author} {\bibfnamefont {D.~N.}\
  \bibnamefont {Christodoulides}},\ and\ \bibinfo {author} {\bibfnamefont
  {M.}~\bibnamefont {Khajavikhan}},\ }\bibfield  {title} {\bibinfo {title}
  {Topological insulator laser: Experiments},\ }\href
  {https://doi.org/10.1126/science.aar4005} {\bibfield  {journal} {\bibinfo
  {journal} {Science}\ }\textbf {\bibinfo {volume} {359}},\ \bibinfo {pages}
  {eaar4005} (\bibinfo {year} {2018})}\BibitemShut {NoStop}%
\bibitem [{\citenamefont {Harari}\ \emph {et~al.}(2018)\citenamefont {Harari},
  \citenamefont {Bandres}, \citenamefont {Lumer}, \citenamefont {Rechtsman},
  \citenamefont {Chong}, \citenamefont {Khajavikhan}, \citenamefont
  {Christodoulides},\ and\ \citenamefont {Segev}}]{Harari2018}%
  \BibitemOpen
  \bibfield  {author} {\bibinfo {author} {\bibfnamefont {G.}~\bibnamefont
  {Harari}}, \bibinfo {author} {\bibfnamefont {M.~A.}\ \bibnamefont {Bandres}},
  \bibinfo {author} {\bibfnamefont {Y.}~\bibnamefont {Lumer}}, \bibinfo
  {author} {\bibfnamefont {M.~C.}\ \bibnamefont {Rechtsman}}, \bibinfo {author}
  {\bibfnamefont {Y.~D.}\ \bibnamefont {Chong}}, \bibinfo {author}
  {\bibfnamefont {M.}~\bibnamefont {Khajavikhan}}, \bibinfo {author}
  {\bibfnamefont {D.~N.}\ \bibnamefont {Christodoulides}},\ and\ \bibinfo
  {author} {\bibfnamefont {M.}~\bibnamefont {Segev}},\ }\bibfield  {title}
  {\bibinfo {title} {Topological insulator laser: Theory},\ }\href
  {https://doi.org/10.1126/science.aar4003} {\bibfield  {journal} {\bibinfo
  {journal} {Science}\ }\textbf {\bibinfo {volume} {359}},\ \bibinfo {pages}
  {eaar4003} (\bibinfo {year} {2018})}\BibitemShut {NoStop}%
\bibitem [{\citenamefont {Longhi}\ \emph {et~al.}(2018)\citenamefont {Longhi},
  \citenamefont {Kominis},\ and\ \citenamefont {Kovanis}}]{Longhi2018}%
  \BibitemOpen
  \bibfield  {author} {\bibinfo {author} {\bibfnamefont {S.}~\bibnamefont
  {Longhi}}, \bibinfo {author} {\bibfnamefont {Y.}~\bibnamefont {Kominis}},\
  and\ \bibinfo {author} {\bibfnamefont {V.}~\bibnamefont {Kovanis}},\
  }\bibfield  {title} {\bibinfo {title} {Presence of temporal dynamical
  instabilities in topological insulator lasers},\ }\href
  {https://doi.org/10.1209/0295-5075/122/14004} {\bibfield  {journal} {\bibinfo
   {journal} {Europhys. Lett.}\ }\textbf {\bibinfo {volume} {122}},\ \bibinfo
  {pages} {14004} (\bibinfo {year} {2018})}\BibitemShut {NoStop}%
\bibitem [{\citenamefont {Parto}\ \emph {et~al.}(2018)\citenamefont {Parto},
  \citenamefont {Wittek}, \citenamefont {Hodaei}, \citenamefont {Harari},
  \citenamefont {Bandres}, \citenamefont {Ren}, \citenamefont {Rechtsman},
  \citenamefont {Segev}, \citenamefont {Christodoulides},\ and\ \citenamefont
  {Khajavikhan}}]{Parto2018}%
  \BibitemOpen
  \bibfield  {author} {\bibinfo {author} {\bibfnamefont {M.}~\bibnamefont
  {Parto}}, \bibinfo {author} {\bibfnamefont {S.}~\bibnamefont {Wittek}},
  \bibinfo {author} {\bibfnamefont {H.}~\bibnamefont {Hodaei}}, \bibinfo
  {author} {\bibfnamefont {G.}~\bibnamefont {Harari}}, \bibinfo {author}
  {\bibfnamefont {M.~A.}\ \bibnamefont {Bandres}}, \bibinfo {author}
  {\bibfnamefont {J.}~\bibnamefont {Ren}}, \bibinfo {author} {\bibfnamefont
  {M.~C.}\ \bibnamefont {Rechtsman}}, \bibinfo {author} {\bibfnamefont
  {M.}~\bibnamefont {Segev}}, \bibinfo {author} {\bibfnamefont {D.~N.}\
  \bibnamefont {Christodoulides}},\ and\ \bibinfo {author} {\bibfnamefont
  {M.}~\bibnamefont {Khajavikhan}},\ }\bibfield  {title} {\bibinfo {title}
  {Edge-mode lasing in 1{D} topological active arrays},\ }\href
  {https://doi.org/10.1103/PhysRevLett.120.113901} {\bibfield  {journal}
  {\bibinfo  {journal} {Phys. Rev. Lett.}\ }\textbf {\bibinfo {volume} {120}},\
  \bibinfo {pages} {113901} (\bibinfo {year} {2018})}\BibitemShut {NoStop}%
\bibitem [{\citenamefont {Zhao}\ \emph {et~al.}(2018)\citenamefont {Zhao},
  \citenamefont {Miao}, \citenamefont {Teimourpour}, \citenamefont {Malzard},
  \citenamefont {El-{G}anainy}, \citenamefont {Schomerus},\ and\ \citenamefont
  {Feng}}]{Zhao2018}%
  \BibitemOpen
  \bibfield  {author} {\bibinfo {author} {\bibfnamefont {H.}~\bibnamefont
  {Zhao}}, \bibinfo {author} {\bibfnamefont {P.}~\bibnamefont {Miao}}, \bibinfo
  {author} {\bibfnamefont {M.~H.}\ \bibnamefont {Teimourpour}}, \bibinfo
  {author} {\bibfnamefont {S.}~\bibnamefont {Malzard}}, \bibinfo {author}
  {\bibfnamefont {R.}~\bibnamefont {El-{G}anainy}}, \bibinfo {author}
  {\bibfnamefont {H.}~\bibnamefont {Schomerus}},\ and\ \bibinfo {author}
  {\bibfnamefont {L.}~\bibnamefont {Feng}},\ }\bibfield  {title} {\bibinfo
  {title} {Topological hybrid silicon microlasers},\ }\href
  {https://doi.org/10.1038/s41467-018-03434-2} {\bibfield  {journal} {\bibinfo
  {journal} {Nat. Commun.}\ }\textbf {\bibinfo {volume} {9}},\ \bibinfo {pages}
  {981} (\bibinfo {year} {2018})}\BibitemShut {NoStop}%
\bibitem [{\citenamefont {Jaramillo-\'Avila}\ \emph {et~al.}(2020)\citenamefont
  {Jaramillo-\'Avila}, \citenamefont {Maldonado-Villamizar},\ and\
  \citenamefont {Rodr\'{\i}guez-Lara}}]{JaramilloAvila2020b}%
  \BibitemOpen
  \bibfield  {author} {\bibinfo {author} {\bibfnamefont {B.}~\bibnamefont
  {Jaramillo-\'Avila}}, \bibinfo {author} {\bibfnamefont {F.~H.}\ \bibnamefont
  {Maldonado-Villamizar}},\ and\ \bibinfo {author} {\bibfnamefont {B.~M.}\
  \bibnamefont {Rodr\'{\i}guez-Lara}},\ }\bibfield  {title} {\bibinfo {title}
  {Optical simulation of atomic decay enhancement and suppression},\ }\href
  {https://doi.org/10.1103/PhysRevA.102.053501} {\bibfield  {journal} {\bibinfo
   {journal} {Phys. Rev. A}\ }\textbf {\bibinfo {volume} {102}},\ \bibinfo
  {pages} {053501} (\bibinfo {year} {2020})}\BibitemShut {NoStop}%
\bibitem [{\citenamefont {Longhi}(2020)}]{Longhi2020}%
  \BibitemOpen
  \bibfield  {author} {\bibinfo {author} {\bibfnamefont {S.}~\bibnamefont
  {Longhi}},\ }\bibfield  {title} {\bibinfo {title} {Photonic simulation of
  giant atom decay},\ }\href {https://doi.org/10.1364/OL.393578} {\bibfield
  {journal} {\bibinfo  {journal} {Opt. Lett.}\ }\textbf {\bibinfo {volume}
  {45}},\ \bibinfo {pages} {3017} (\bibinfo {year} {2020})}\BibitemShut
  {NoStop}%
\bibitem [{\citenamefont {Bender}\ and\ \citenamefont
  {Boettcher}(1998)}]{Bender1998}%
  \BibitemOpen
  \bibfield  {author} {\bibinfo {author} {\bibfnamefont {C.~M.}\ \bibnamefont
  {Bender}}\ and\ \bibinfo {author} {\bibfnamefont {S.}~\bibnamefont
  {Boettcher}},\ }\bibfield  {title} {\bibinfo {title} {Real spectra in
  non-{H}ermitian {H}amiltonians having {$\mathcal{P}\mathcal{T}$} symmetry},\
  }\href {https://doi.org/10.1103/PhysRevLett.80.5243} {\bibfield  {journal}
  {\bibinfo  {journal} {Phys. Rev. Lett.}\ }\textbf {\bibinfo {volume} {80}},\
  \bibinfo {pages} {5243} (\bibinfo {year} {1998})}\BibitemShut {NoStop}%
\bibitem [{\citenamefont {El-Ganainy}\ \emph {et~al.}(2007)\citenamefont
  {El-Ganainy}, \citenamefont {Makris}, \citenamefont {Christodoulides},\ and\
  \citenamefont {Musslimani}}]{ElGanainy2007}%
  \BibitemOpen
  \bibfield  {author} {\bibinfo {author} {\bibfnamefont {R.}~\bibnamefont
  {El-Ganainy}}, \bibinfo {author} {\bibfnamefont {K.~G.}\ \bibnamefont
  {Makris}}, \bibinfo {author} {\bibfnamefont {D.~N.}\ \bibnamefont
  {Christodoulides}},\ and\ \bibinfo {author} {\bibfnamefont {Z.~H.}\
  \bibnamefont {Musslimani}},\ }\bibfield  {title} {\bibinfo {title} {Theory of
  coupled optical {PT}-symmetric structures},\ }\href
  {https://doi.org/10.1364/OL.32.002632} {\bibfield  {journal} {\bibinfo
  {journal} {Opt. Lett.}\ }\textbf {\bibinfo {volume} {32}},\ \bibinfo {pages}
  {2632} (\bibinfo {year} {2007})}\BibitemShut {NoStop}%
\bibitem [{\citenamefont {R{\"u}ter}\ \emph {et~al.}(2010)\citenamefont
  {R{\"u}ter}, \citenamefont {Makris}, \citenamefont {El-Ganainy},
  \citenamefont {Christodoulides}, \citenamefont {Segev},\ and\ \citenamefont
  {Kip}}]{Ruter2010}%
  \BibitemOpen
  \bibfield  {author} {\bibinfo {author} {\bibfnamefont {C.~E.}\ \bibnamefont
  {R{\"u}ter}}, \bibinfo {author} {\bibfnamefont {K.~G.}\ \bibnamefont
  {Makris}}, \bibinfo {author} {\bibfnamefont {R.}~\bibnamefont {El-Ganainy}},
  \bibinfo {author} {\bibfnamefont {D.~N.}\ \bibnamefont {Christodoulides}},
  \bibinfo {author} {\bibfnamefont {M.}~\bibnamefont {Segev}},\ and\ \bibinfo
  {author} {\bibfnamefont {D.}~\bibnamefont {Kip}},\ }\bibfield  {title}
  {\bibinfo {title} {Observation of parity–time symmetry in optics},\ }\href
  {https://doi.org/10.1038/nphys1515} {\bibfield  {journal} {\bibinfo
  {journal} {Nat. Phys.}\ }\textbf {\bibinfo {volume} {32}},\ \bibinfo {pages}
  {6} (\bibinfo {year} {2010})}\BibitemShut {NoStop}%
\bibitem [{\citenamefont {{Huerta Morales}}\ \emph {et~al.}(2016)\citenamefont
  {{Huerta Morales}}, \citenamefont {Guerrero}, \citenamefont
  {L{\'o}pez-Aguayo},\ and\ \citenamefont
  {Rodr{\'i}guez-Lara}}]{HuertaMorales2016}%
  \BibitemOpen
  \bibfield  {author} {\bibinfo {author} {\bibfnamefont {J.~D.}\ \bibnamefont
  {{Huerta Morales}}}, \bibinfo {author} {\bibfnamefont {J.}~\bibnamefont
  {Guerrero}}, \bibinfo {author} {\bibfnamefont {S.}~\bibnamefont
  {L{\'o}pez-Aguayo}},\ and\ \bibinfo {author} {\bibfnamefont {B.~M.}\
  \bibnamefont {Rodr{\'i}guez-Lara}},\ }\bibfield  {title} {\bibinfo {title}
  {Revisiting the optical {$\mathcal{PT}$}-symmetric dimer},\ }\href
  {https://doi.org/10.3390/sym8090083} {\bibfield  {journal} {\bibinfo
  {journal} {Symmetry}\ }\textbf {\bibinfo {volume} {8}},\ \bibinfo {pages}
  {83} (\bibinfo {year} {2016})}\BibitemShut {NoStop}%
\bibitem [{\citenamefont {Hodaei}\ \emph {et~al.}(2014)\citenamefont {Hodaei},
  \citenamefont {Miri}, \citenamefont {Heinrich}, \citenamefont
  {Christodoulides},\ and\ \citenamefont {Khajavikhan}}]{Hodaei2014}%
  \BibitemOpen
  \bibfield  {author} {\bibinfo {author} {\bibfnamefont {H.}~\bibnamefont
  {Hodaei}}, \bibinfo {author} {\bibfnamefont {M.-A.}\ \bibnamefont {Miri}},
  \bibinfo {author} {\bibfnamefont {M.}~\bibnamefont {Heinrich}}, \bibinfo
  {author} {\bibfnamefont {D.~N.}\ \bibnamefont {Christodoulides}},\ and\
  \bibinfo {author} {\bibfnamefont {M.}~\bibnamefont {Khajavikhan}},\
  }\bibfield  {title} {\bibinfo {title} {Parity-time–symmetric microring
  lasers},\ }\href {https://doi.org/10.1126/science.1258480} {\bibfield
  {journal} {\bibinfo  {journal} {Science}\ }\textbf {\bibinfo {volume}
  {346}},\ \bibinfo {pages} {975} (\bibinfo {year} {2014})}\BibitemShut
  {NoStop}%
\bibitem [{\citenamefont {Hodaei}\ \emph {et~al.}(2016)\citenamefont {Hodaei},
  \citenamefont {Miri}, \citenamefont {Hassan}, \citenamefont {Hayenga},
  \citenamefont {Heinrich}, \citenamefont {Christodoulides},\ and\
  \citenamefont {Khajavikhan}}]{Hodaei2016}%
  \BibitemOpen
  \bibfield  {author} {\bibinfo {author} {\bibfnamefont {H.}~\bibnamefont
  {Hodaei}}, \bibinfo {author} {\bibfnamefont {M.-A.}\ \bibnamefont {Miri}},
  \bibinfo {author} {\bibfnamefont {A.~U.}\ \bibnamefont {Hassan}}, \bibinfo
  {author} {\bibfnamefont {W.~E.}\ \bibnamefont {Hayenga}}, \bibinfo {author}
  {\bibfnamefont {M.}~\bibnamefont {Heinrich}}, \bibinfo {author}
  {\bibfnamefont {D.~N.}\ \bibnamefont {Christodoulides}},\ and\ \bibinfo
  {author} {\bibfnamefont {M.}~\bibnamefont {Khajavikhan}},\ }\bibfield
  {title} {\bibinfo {title} {Single mode lasing in transversely multi-moded
  pt-symmetric microring resonators},\ }\href
  {https://doi.org/10.1002/lpor.201500292} {\bibfield  {journal} {\bibinfo
  {journal} {Laser Photonic. Rev.}\ }\textbf {\bibinfo {volume} {10}},\
  \bibinfo {pages} {494} (\bibinfo {year} {2016})}\BibitemShut {NoStop}%
\bibitem [{\citenamefont {Ren}\ \emph {et~al.}(2018)\citenamefont {Ren},
  \citenamefont {Liu}, \citenamefont {Parto}, \citenamefont {Hayenga},
  \citenamefont {Hokmabadi}, \citenamefont {Christodoulides},\ and\
  \citenamefont {Khajavikhan}}]{Ren2018}%
  \BibitemOpen
  \bibfield  {author} {\bibinfo {author} {\bibfnamefont {J.}~\bibnamefont
  {Ren}}, \bibinfo {author} {\bibfnamefont {Y.~G.~N.}\ \bibnamefont {Liu}},
  \bibinfo {author} {\bibfnamefont {M.}~\bibnamefont {Parto}}, \bibinfo
  {author} {\bibfnamefont {W.~E.}\ \bibnamefont {Hayenga}}, \bibinfo {author}
  {\bibfnamefont {M.~P.}\ \bibnamefont {Hokmabadi}}, \bibinfo {author}
  {\bibfnamefont {D.~N.}\ \bibnamefont {Christodoulides}},\ and\ \bibinfo
  {author} {\bibfnamefont {M.}~\bibnamefont {Khajavikhan}},\ }\bibfield
  {title} {\bibinfo {title} {Unidirectional light emission in {PT}-symmetric
  microring lasers},\ }\href {https://doi.org/10.1364/OE.26.027153} {\bibfield
  {journal} {\bibinfo  {journal} {Opt. Express}\ }\textbf {\bibinfo {volume}
  {26}},\ \bibinfo {pages} {27153} (\bibinfo {year} {2018})}\BibitemShut
  {NoStop}%
\bibitem [{\citenamefont {Zyablovsky}\ \emph
  {et~al.}(2014{\natexlab{a}})\citenamefont {Zyablovsky}, \citenamefont
  {Vinogradov}, \citenamefont {Pukhov}, \citenamefont {Dorofeenko},\ and\
  \citenamefont {Lisyansky}}]{Zyablovsky2014a}%
  \BibitemOpen
  \bibfield  {author} {\bibinfo {author} {\bibfnamefont {A.~A.}\ \bibnamefont
  {Zyablovsky}}, \bibinfo {author} {\bibfnamefont {A.~P.}\ \bibnamefont
  {Vinogradov}}, \bibinfo {author} {\bibfnamefont {A.~A.}\ \bibnamefont
  {Pukhov}}, \bibinfo {author} {\bibfnamefont {A.~V.}\ \bibnamefont
  {Dorofeenko}},\ and\ \bibinfo {author} {\bibfnamefont {A.~A.}\ \bibnamefont
  {Lisyansky}},\ }\bibfield  {title} {\bibinfo {title} {{PT}-symmetry in
  optics},\ }\href {https://doi.org/10.3367/UFNe.0184.201411b.1177} {\bibfield
  {journal} {\bibinfo  {journal} {Phys.-Usp.}\ }\textbf {\bibinfo {volume}
  {57}},\ \bibinfo {pages} {1063} (\bibinfo {year}
  {2014}{\natexlab{a}})}\BibitemShut {NoStop}%
\bibitem [{\citenamefont {Zyablovsky}\ \emph
  {et~al.}(2014{\natexlab{b}})\citenamefont {Zyablovsky}, \citenamefont
  {Vinogradov}, \citenamefont {Dorofeenko}, \citenamefont {Pukhov},\ and\
  \citenamefont {Lisyansky}}]{Zyablovsky2014b}%
  \BibitemOpen
  \bibfield  {author} {\bibinfo {author} {\bibfnamefont {A.~A.}\ \bibnamefont
  {Zyablovsky}}, \bibinfo {author} {\bibfnamefont {A.~P.}\ \bibnamefont
  {Vinogradov}}, \bibinfo {author} {\bibfnamefont {A.~V.}\ \bibnamefont
  {Dorofeenko}}, \bibinfo {author} {\bibfnamefont {A.~A.}\ \bibnamefont
  {Pukhov}},\ and\ \bibinfo {author} {\bibfnamefont {A.~A.}\ \bibnamefont
  {Lisyansky}},\ }\bibfield  {title} {\bibinfo {title} {Causality and phase
  transitions in $\mathcal{PT}$-symmetric optical systems},\ }\href
  {https://doi.org/10.1103/PhysRevA.89.033808} {\bibfield  {journal} {\bibinfo
  {journal} {Phys. Rev. A}\ }\textbf {\bibinfo {volume} {89}},\ \bibinfo
  {pages} {033808} (\bibinfo {year} {2014}{\natexlab{b}})}\BibitemShut
  {NoStop}%
\bibitem [{\citenamefont {Baili}\ \emph {et~al.}(2009)\citenamefont {Baili},
  \citenamefont {Alouini}, \citenamefont {Malherbe}, \citenamefont {Dolfi},
  \citenamefont {Sagnes},\ and\ \citenamefont {Bretenaker}}]{Baili2009}%
  \BibitemOpen
  \bibfield  {author} {\bibinfo {author} {\bibfnamefont {G.}~\bibnamefont
  {Baili}}, \bibinfo {author} {\bibfnamefont {M.}~\bibnamefont {Alouini}},
  \bibinfo {author} {\bibfnamefont {T.}~\bibnamefont {Malherbe}}, \bibinfo
  {author} {\bibfnamefont {D.}~\bibnamefont {Dolfi}}, \bibinfo {author}
  {\bibfnamefont {I.}~\bibnamefont {Sagnes}},\ and\ \bibinfo {author}
  {\bibfnamefont {F.}~\bibnamefont {Bretenaker}},\ }\bibfield  {title}
  {\bibinfo {title} {Direct observation of the class-{B} to class-{A}
  transition in the dynamical behavior of a semiconductor laser},\ }\href
  {https://doi.org/10.1209/0295-5075/87/44005} {\bibfield  {journal} {\bibinfo
  {journal} {Europhys. Lett.}\ }\textbf {\bibinfo {volume} {87}},\ \bibinfo
  {pages} {44005} (\bibinfo {year} {2009})}\BibitemShut {NoStop}%
\bibitem [{\citenamefont {Ohtsubo}(2017)}]{Ohtsubo2017}%
  \BibitemOpen
  \bibinfo {editor} {\bibfnamefont {J.}~\bibnamefont {Ohtsubo}},\ ed.,\ \href
  {https://doi.org/10.1007/978-3-319-56138-7} {\emph {\bibinfo {title}
  {Semiconductor Lasers}}}\ (\bibinfo  {publisher} {Springer},\ \bibinfo {year}
  {2017})\BibitemShut {NoStop}%
\bibitem [{\citenamefont {{G{\"u}emes Frese}}\ \emph
  {et~al.}(2022)\citenamefont {{G{\"u}emes Frese}}, \citenamefont
  {Jaramillo-{\'A}vila},\ and\ \citenamefont
  {Rodr{\'i}guez-Lara}}]{GuemesFrese2022}%
  \BibitemOpen
  \bibfield  {author} {\bibinfo {author} {\bibfnamefont {L.~E.}\ \bibnamefont
  {{G{\"u}emes Frese}}}, \bibinfo {author} {\bibfnamefont {B.~R.}\ \bibnamefont
  {Jaramillo-{\'A}vila}},\ and\ \bibinfo {author} {\bibfnamefont {B.~M.}\
  \bibnamefont {Rodr{\'i}guez-Lara}},\ }\bibfield  {title} {\bibinfo {title}
  {Routes to chaos in a class-{B} laser coupled to a neutral resonator},\
  }\href {https://doi.org/10.1103/PhysRevA.106.033507} {\bibfield  {journal}
  {\bibinfo  {journal} {Phys. Rev. A}\ }\textbf {\bibinfo {volume} {106}},\
  \bibinfo {pages} {033507} (\bibinfo {year} {2022})}\BibitemShut {NoStop}%
\bibitem [{\citenamefont {{Nodal Stevens}}\ \emph {et~al.}(2018)\citenamefont
  {{Nodal Stevens}}, \citenamefont {Jaramillo-{\'A}vila},\ and\ \citenamefont
  {Rodr{\'i}guez-Lara}}]{NodalStevens2018}%
  \BibitemOpen
  \bibfield  {author} {\bibinfo {author} {\bibfnamefont {D.~J.}\ \bibnamefont
  {{Nodal Stevens}}}, \bibinfo {author} {\bibfnamefont {B.~R.}\ \bibnamefont
  {Jaramillo-{\'A}vila}},\ and\ \bibinfo {author} {\bibfnamefont {B.~M.}\
  \bibnamefont {Rodr{\'i}guez-Lara}},\ }\bibfield  {title} {\bibinfo {title}
  {Necklaces of $\mathcal{PT}$-symmetric dimers},\ }\href
  {https://doi.org/10.1364/PRJ.6.000A31} {\bibfield  {journal} {\bibinfo
  {journal} {Photon. Res.}\ }\textbf {\bibinfo {volume} {6}},\ \bibinfo {pages}
  {A31} (\bibinfo {year} {2018})}\BibitemShut {NoStop}%
\bibitem [{\citenamefont {Rodr\'{i}guez-Lara}\ and\ \citenamefont
  {Guerrero}(2015)}]{RodriguezLara2015}%
  \BibitemOpen
  \bibfield  {author} {\bibinfo {author} {\bibfnamefont {B.~M.}\ \bibnamefont
  {Rodr\'{i}guez-Lara}}\ and\ \bibinfo {author} {\bibfnamefont
  {J.}~\bibnamefont {Guerrero}},\ }\bibfield  {title} {\bibinfo {title}
  {Optical finite representation of the {L}orentz group},\ }\href
  {https://doi.org/10.1364/OL.40.005682} {\bibfield  {journal} {\bibinfo
  {journal} {Opt. Lett.}\ }\textbf {\bibinfo {volume} {40}},\ \bibinfo {pages}
  {5682} (\bibinfo {year} {2015})}\BibitemShut {NoStop}%
\bibitem [{\citenamefont {Chen}\ and\ \citenamefont {Jung}(2016)}]{Chen2016}%
  \BibitemOpen
  \bibfield  {author} {\bibinfo {author} {\bibfnamefont {P.-Y.}\ \bibnamefont
  {Chen}}\ and\ \bibinfo {author} {\bibfnamefont {J.}~\bibnamefont {Jung}},\
  }\bibfield  {title} {\bibinfo {title} {$\mathcal{P}\mathcal{T}$ symmetry and
  singularity-enhanced sensing based on photoexcited graphene metasurfaces},\
  }\href {https://doi.org/10.1103/PhysRevApplied.5.064018} {\bibfield
  {journal} {\bibinfo  {journal} {Phys. Rev. Appl.}\ }\textbf {\bibinfo
  {volume} {5}},\ \bibinfo {pages} {064018} (\bibinfo {year}
  {2016})}\BibitemShut {NoStop}%
\bibitem [{\citenamefont {Arecchi}\ \emph {et~al.}(1984)\citenamefont
  {Arecchi}, \citenamefont {Lippi}, \citenamefont {Puccioni},\ and\
  \citenamefont {Tredicce}}]{Arecchi1984}%
  \BibitemOpen
  \bibfield  {author} {\bibinfo {author} {\bibfnamefont {F.~T.}\ \bibnamefont
  {Arecchi}}, \bibinfo {author} {\bibfnamefont {G.~L.}\ \bibnamefont {Lippi}},
  \bibinfo {author} {\bibfnamefont {G.~P.}\ \bibnamefont {Puccioni}},\ and\
  \bibinfo {author} {\bibfnamefont {J.~R.}\ \bibnamefont {Tredicce}},\
  }\bibfield  {title} {\bibinfo {title} {Deterministic chaos in laser with
  injected signal},\ }\href {https://doi.org/10.1016/0030-4018(84)90016-6}
  {\bibfield  {journal} {\bibinfo  {journal} {Opt. Commun.}\ }\textbf {\bibinfo
  {volume} {51}},\ \bibinfo {pages} {308} (\bibinfo {year} {1984})}\BibitemShut
  {NoStop}%
\bibitem [{\citenamefont {Wieczorek}\ \emph {et~al.}(2005)\citenamefont
  {Wieczorek}, \citenamefont {Krauskropf}, \citenamefont {Simpson},\ and\
  \citenamefont {Lenstra}}]{Wieczorek2005}%
  \BibitemOpen
  \bibfield  {author} {\bibinfo {author} {\bibfnamefont {S.}~\bibnamefont
  {Wieczorek}}, \bibinfo {author} {\bibfnamefont {B.}~\bibnamefont
  {Krauskropf}}, \bibinfo {author} {\bibfnamefont {T.~B.}\ \bibnamefont
  {Simpson}},\ and\ \bibinfo {author} {\bibfnamefont {D.}~\bibnamefont
  {Lenstra}},\ }\bibfield  {title} {\bibinfo {title} {The dynamical complexity
  of optical injected semiconductor lasers},\ }\href
  {https://doi.org/10.1016/j.physrep.2005.06.003} {\bibfield  {journal}
  {\bibinfo  {journal} {Phys. Rep.}\ }\textbf {\bibinfo {volume} {416}},\
  \bibinfo {pages} {1} (\bibinfo {year} {2005})}\BibitemShut {NoStop}%
\bibitem [{\citenamefont {Guckenheimer}\ and\ \citenamefont
  {Holmes}(1983)}]{Guckenheimer1983}%
  \BibitemOpen
  \bibfield  {author} {\bibinfo {author} {\bibfnamefont {J.}~\bibnamefont
  {Guckenheimer}}\ and\ \bibinfo {author} {\bibfnamefont {P.}~\bibnamefont
  {Holmes}},\ }\href {https://doi.org/10.1007/978-1-4612-1140-2} {\emph
  {\bibinfo {title} {Nonlinear Ordinary Differential Equations}}}\ (\bibinfo
  {publisher} {Springer},\ \bibinfo {address} {New York},\ \bibinfo {year}
  {1983})\BibitemShut {NoStop}%
\bibitem [{\citenamefont {Murray}(1989)}]{Murray1989}%
  \BibitemOpen
  \bibfield  {author} {\bibinfo {author} {\bibfnamefont {J.~D.}\ \bibnamefont
  {Murray}},\ }\href {https://doi.org/10.1007/b98868} {\emph {\bibinfo {title}
  {Mathematical Biology}}}\ (\bibinfo  {publisher} {Springer},\ \bibinfo
  {address} {Berlin},\ \bibinfo {year} {1989})\BibitemShut {NoStop}%
\bibitem [{\citenamefont {Jordan}\ and\ \citenamefont
  {Smith}(2007)}]{Jordan2007}%
  \BibitemOpen
  \bibfield  {author} {\bibinfo {author} {\bibfnamefont {D.}~\bibnamefont
  {Jordan}}\ and\ \bibinfo {author} {\bibfnamefont {P.}~\bibnamefont {Smith}},\
  }\href@noop {} {\emph {\bibinfo {title} {Nonlinear Ordinary Differential
  Equations}}}\ (\bibinfo  {publisher} {Oxford University Press},\ \bibinfo
  {address} {Oxford},\ \bibinfo {year} {2007})\BibitemShut {NoStop}%
\end{thebibliography}

%

\end{document}